\documentclass[12pt]{article}

\usepackage{amsmath}
\usepackage{amssymb}

\advance\hoffset by -7mm   
\makeatletter 
\@addtoreset{equation}{section}  
\makeatother 
 
\def\ii{\'\i}

\def\ftoday{{\sl {Le \number\day \space\ifcase\month 
\or janvier\or f\'evrier\or mars\or avril\or mai
\or juin\or juillet\or ao\^ut\or septembre\or octobre
\or novembre \or d\'ecembre\fi\space \number\year}}}    
\def\ptoday{{\sl {\number\day \space de\space \ifcase\month 
\or janeiro\or fevereiro\or mar{\c c}o\or abril\or maio
\or junho\or julho\or agosto\or setembro\or outubro
\or novembro \or dezembro\fi\space de\space \number\year}}}    
\def\gtoday{{\sl {Den \number\day. \ifcase\month 
\or Januar\or Februar\or M\"arz\or April\or Mai
\or Juni\or Juli\or August\or September\or Oktober
\or November \or Dezember\fi\space \number\year}}}    
\def\today{{\sl {\ifcase\month
\or January\or February\or March\or April\or May
\or June\or July\or August\or September\or October
\or November \or December\fi \space\number\day,\space 
                                            \number\year}}}
\newcommand{\journal}[4]{{\em #1~}#2\,(#3)\,#4}

\newcommand{\ijmp}{\journal {Int. J. Mod. Phys.}}

\newcommand{\cqg}{\journal {Class. Quantum Grav.}}

\newcommand{\np}{\journal {Nucl. Phys.}}

\newcommand{\ptp}{\journal {Progr. Theor. Phys.}}

\newcommand{\annp}{\journal {Ann. Phys. (N.Y.)}}

\renewcommand{\a}{\alpha}
\renewcommand{\b}{\beta}
\newcommand{\g}{\gamma}           \newcommand{\GA}{\Gamma}
\renewcommand{\d}{\delta}         
\newcommand{\e}{\varepsilon}
\newcommand{\ka}{\kappa}
        \newcommand{\LA}{\Lambda}
\newcommand{\m}{\mu}
\newcommand{\n}{\nu}
\newcommand{\om}{\omega}         \newcommand{\OM}{\Omega}
\newcommand{\s}{\sigma}           \renewcommand{\S}{\Sigma}
           \newcommand{\F}{{\Phi}}

\renewcommand{\AA}{{\cal A}}

\newcommand{\FF}{{\cal F}}
\newcommand{\GG}{{\cal G}}
\newcommand{\HH}{{\cal H}}

\newcommand{\XX}{{\cal X}}

\newcommand{\esp}{\\[3mm]}
\newcommand{\ES}{\\[6mm]}

\newcommand{\sla}{\raise.15ex\hbox{$/$}\kern -.57em} 
\newcommand{\Sla}{\raise.15ex\hbox{$/$}\kern -.70em}

\newcommand{\lp}{\left(}\newcommand{\rp}{\right)}

\newcommand{\lac}{\left\{}

\newcommand{\complex}{{\kern .1em {\raise .47ex
\hbox {$\scriptscriptstyle |$}}
    \kern -.4em {\rm C}}}
\newcommand{\real}{{{\rm I} \kern -.19em {\rm R}}}
\newcommand{\rational}{{\kern .1em {\raise .47ex
\hbox{$\scripscriptstyle |$}}
    \kern -.35em {\rm Q}}}
\renewcommand{\natural}{{\vrule height 1.6ex width
.05em depth 0ex \kern -.35em {\rm N}}}

\newcommand{\pa}{\partial}

\renewcommand{\dfrac}[2]{{\displaystyle{\frac{#1}{#2}}}}
\newcommand{\dsum}[2]{\displaystyle{\sum_{#1}^{#2}}}   
\newcommand{\dint}{\displaystyle{\int}}
\newcommand{\eg}{{\em e.g.,\ }}

\newcommand{\ie}{{{\em i.e.},\ }}

\newcommand{\twiddle}{\lower.9ex\rlap{$\kern -.1em\scriptstyle\sim$}}

\newcommand{\bra}[1]{\left\langle {#1}\right|}
\newcommand{\ket}[1]{\left| {#1}\right\rangle}

\newcommand{\vev}[1]{\langle {#1}\rangle}

\newcommand{\equ}[1]{(\ref{#1})}
\newcommand{\eq}{\begin{equation}}
\newcommand{\eqn}[1]{\label{#1}\end{equation}}
\newcommand{\eea}{\end{eqnarray}}
\newcommand{\eqa}{\begin{eqnarray}}
\newcommand{\eqan}[1]{\label{#1}\end{eqnarray}}
\newcommand{\ba}{\begin{array}}
\newcommand{\ea}{\end{array}}
\newcommand{\eqac}{\begin{equation}\begin{array}{rcl}}
\newcommand{\eqacn}[1]{\end{array}\label{#1}\end{equation}}

\newcommand{\bz}{\begin{enumerate}}
\newcommand{\ez}{\end{enumerate}}

\newcommand{\bfe}{{\boldsymbol{e}}}
\newcommand{\bfom}{\boldsymbol{\omega}}
\newcommand{\bfT}{{\boldsymbol{T}}}

\newcommand{\bfA}{{\boldsymbol{A}}} 
\newcommand{\bfB}{{\boldsymbol{B}}}
\newcommand{\bfD}{{\mathbf{D}}}
\newcommand{\bfF}{{\boldsymbol{F}}}
\newcommand{\bfR}{{\boldsymbol{R}}}
\newcommand{\bfd}{{\mathbf{d}}}
\newcommand{\bfx}{{\boldsymbol{x}}}
\newcommand{\bfy}{{\boldsymbol{y}}}



\begin{document}
\global\long\def\eb{\mathbf{e}}

\global\long\def\A{\boldsymbol{\mathsf{A}}}

\global\long\def\B{\boldsymbol{\mathsf{B}}}

\global\long\def\F{\boldsymbol{\mathsf{F}}}

\global\long\def\R{\mathsf{R}}

\global\long\def\T{\mathsf{T}}

\global\long\def\M{\mathcal{M}}

\global\long\def\V{\mathbb{V}}




\global\long\def\eps{\boldsymbol{\varepsilon}}

\global\long\def\bOM{\boldsymbol{\Omega}}

\global\long\def\Vol{\mathrm{Vol}}

\global\long\def\so{\mathfrak{so}(3,1)}

\global\long\def\tr{\mathrm{tr}}

\global\long\def\im{\mathrm{i}}

\global\long\def\pd#1#2{\frac{\partial#2}{\partial#1}}

\global\long\def\td#1#2{\frac{d#2}{d#1}}

\global\long\def\ded#1#2#3{\frac{\delta#3}{\delta A_{\theta}^{#1}#2}}

\global\long\def\inp#1#2{\langle#1\,,#2\rangle}

\global\long\def\is{\!}

\global\long\def\udd#1#2#3{#1^{#2}\is_{#3}}

\global\long\def\dud#1#2#3#4{#1_{#2}\!^{#3}\!_{#4}}

\global\long\def\duu#1#2#3{#1_{#2}\is^{#3}}

\global\long\def\bra#1{\langle#1|}

\global\long\def\ket#1{|#1\rangle}

\title{Quantization of Lorentzian 3d Gravity\\  by Partial Gauge Fixing}

\author{Rodrigo M S Barbosa\footnote{Work supported
   in part by the Conselho Nacional de Desenvolvimento Cient\'{\i}fico e
   Tecnol\'{o}gico -- CNPq (Brazil) and 
by the PRONEX project No. 35885149/2006 from FAPES -- CNPq (Brazil).}\ \footnote{Work supported by the Conselho Nacional de Desenvolvimento 
Cient\'{\i}fico e Tecnol\'ologico -- CNPq 
(Brazil) grant no. 130253/2009-0.}\ , 
Clisthenis P Constantinidis$^{*}$,\\
Zui Oporto$^{*}$\footnote{Work supported
   in part by the Centro Latino-Americano de F\ii sica -- CLAF and the Conselho Nacional de Desenvolvimento Cient\'{\i}fico e
   Tecnol\'{o}gico -- CNPq (Brazil) grant No. 141579/2008-0.}\ 
and Olivier Piguet$^{*}$\footnote{Present address: Departamento de F\ii sica, 
Universidade Federal de Vi\c cosa -- UFV, Vi\c cosa, MG, Brazil}\\
$\ $\\
{\small Departamento de F\ii sica, Universidade Federal do Esp\'{\i}rito Santo (UFES)}\\
{\small Vit\'oria, ES, Brazil}
}

\date{}

\maketitle

\begin{center} 

\vspace{-5mm}

{\small\tt 
E-mails: rodrigo\_martins@email.com,
cpconstantinidis@pq.cnpq.br, 
azurnasirpal@gmail.com,
opiguet@yahoo.com}
\end{center}

\vspace{5mm}

\begin{abstract}
$D=2+1$ gravity with a cosmological constant $\LA$ has been shown by
Bonzom and Livine to present a Barbero-Immirzi like ambiguity depending
on a parameter $\g$. We make use of this fact to show that, for $\LA$
$>0$, the Lorentzian theory can be partially gauge fixed and reduced to
an SU(2) Chern-Simons theory. We then review  the already
known quantization of the latter in the
framework of Loop Quantization for the case of space being 
topogically a cylinder. 
We finally construct, in the same setting, a quantum observable which,
although non-trivial at the quantum level, corresponds to a null
classical quantity.

\end{abstract}

\section{Introduction}

Since the discovery by Achucarro and Townsend ~\cite{Achucarro:1987vz}
and the elaborated work of Witten \cite{Witten:1988hc}, it is a sort
of common sense to affirm that 3d gravity and Chern-Simons (CS) gauge
theory are equivalent, up to a total derivative (boundary terms),
with the Poincar\'e group as the underlying gauge group. However, the
difference between the two theories is that in 3d gravity the triad
is restricted to being invertible, whereas no such restriction exists
in CS theory. 
Thus, we can think about CS theory as an extension of
3d gravity including singular metrics or, alternatively, think about
3d gravity as a restricted version of CS theory~\cite{Witten:1988hc}. 
Questions regarding
the role of non-invertible triad is far from being trivial \cite{Matschull:1999he}.

A richer structure emerges by enlarging the local symmetry to the
(anti) de Sitter group. In this case, besides the standard action
for 3d gravity, it is possible to construct an ``exotic''
action equivalent to the former at the level of classical field equations.
This peculiarity of CS gravity was not unnoticed in original Witten's
paper.
This leads to an analogy with a well established
(although still controversial) feature of Loop Quantum Gravity (LQG) known as
the Barbero-Immirzi parameter ambiguity~\cite{general-ref,barbero-immirzi}.
This analogy was studied in detail in~\cite{bonzom-livine},
and represents
the principal motivation of the present work.

The introduction of the  Ashtekar variables within the formalism of
canonical gravity in 4d space-time~\cite{ashtekar,general-ref,nicolai-etc} 
was a big step into the simplification of the constraints.
The simplicity was achieved thanks to the introduction of a complex phase-space,
that turns out to be problematic at the moment of imposing reality conditions.
A self-dual Lagrangian density corresponding to Ashtekar's
Hamiltonian was given independently by Samuel~\cite{samuel} and Jacobson
and Smolin~\cite{jacobson-smolin}. 
Then Barbero pointed out~\cite{barbero-immirzi} the 
possibility of a real realization of canonical variables closely related to Ashtekar's,
but the price to be paid is the introduction of an arbitrary real/complex
parameter $\gamma$ known as the Barbero-Immirzi parameter). Also
the simplicity of Ashtekar's Hamiltonian constraints is destroyed.
Finally Holst introduced~\cite{holst} the most general action  integrating
all the previous cases, which contains a (on-shell) topological
term coupled via the $\gamma$ parameter. The theory is then interpreted
as a class of actions for gravity classically indistinguishable but
quantum mechanically inequivalent.
 Although the introduction of 
 Ashtekar variables fits naturally in 2+1 gravity, it remained up to now 
unknown if it exists a corresponding Ashtekar-Barbero like real connection 
defined on a compact Lie group, e.g. SU(2).

 Bonzom and Livine~\cite{bonzom-livine} proposed a quantization of
Euclidean 3d gravity with a cosmological constant 
based on the presence of two invariant actions,
hence of two independent parameters: the gravitation constant and  a
Barbero-Immirzi like parameter.  In the present paper we  elaborate on this model, 
considering as well
Lorentzian 3d gravity with a cosmological constant $\LA$. In the case
of a positive $\LA$, where the gauge symmetry is de Sitter SO(3,1), we
succeed to quantize the theory in the LQG framework thanks to a gauge
fixing which reduces the gauge invariance group to the compact SO(3)
 or SU(2),  for which a LQG-like quantization
of SU(2) CS theory was developed 
in~\cite{Constantinidis-Luchini-Piguet}, in the case of a 
space manifold with the topology of a cylinder.
We also construct a physical observable which has a certain analogy 
with the area operator of $D=3+1$ 
gravity, although the former does not correspond to any classical geometrical quantity.

 In the LQG scheme, the compactness of the gauge group -- or of the
residual gauge group after a partial gauge fixing -- is known to be
crucial, at least up to now. Apart from Riemannian gravity, this is achieved 
for Lorentzian $D=3+1$ gravity in the time gauge~\cite{general-ref}, which holds 
only in four dimensions. An
important development was brought by the authors of~\cite{thiemann-D>4},
who were able to reduce the gauge group of Lorentzian gravity
 in arbitrary dimension $d+1$ to a compact group SO(d) -- but this works only 
in the Hamiltonian formalism. Our case is
different in three aspects: in three dimensions one does not need their simplicity
constraints (see the second paper of~\cite{thiemann-D>4}), we
rely strongly on the existence of the Barbero-Immirzi like parameter of
Bonzom and Livine, a feature very peculiar to that dimension, and
moreover we start from an existing Lagrangian formalism.

The plan of the paper is as follows. In the next section, which is
mainly a review of results from~\cite{bonzom-livine}, the basic
tools to cast 3d gravity, with (or without) cosmological constant, as
a Chern-Simons gauge theory will be presented, and the appearance
of the Barbero-Immirzi like parameter will be explained. 
In Sections \ref{lambda-positivo} and \ref{calibre-axial} the
main argument of the paper will be presented, in which we consider
the positive cosmological constant model and its gauge fixing. 
The detailed canonical analysis
of  the constrained theory will be developed. 
The construction of an observable and the computation of its spectrum 
are presented in Section \ref{observavel},  where also its classical
counterpart is discussed.
 The last section is reserved
for some comments about the gauge fixing reduction and quantization of
the model in the spirit of LQG, as well as for the conclusions of the
work.

\section{Gravity from Chern-Simons theory with \\ Barbero-Immirzi ambiguity.}
\label{revisao}

To begin with, let $\mathcal{M}$ be an orientable three-dimensional
manifold. In addition, let $G$ be the gauge group and $\mathfrak{g}$
its Lie algebra equipped with a non-degenerate invariant quadratic
form $\langle\cdot,\cdot\rangle$.

The CS action is defined as\footnote{We don't write explicitly the wedge symbol
$\wedge$ for the external product of forms.}  
\begin{equation}
S=-\frac{\kappa}{2}\int_{\mathcal{M}}\langle\Omega,\mathrm{d}\Omega
+\frac{2}{3}\Omega\, \Omega\rangle.\label{CS}
\end{equation}
 where $\Omega=\Omega_{\mu}\mathrm{d}x^{\mu}$ is a $\mathfrak{g}$-valued
1-form connection and $\kappa$ a dimensionless constant\footnote{In what follows 
Greek indexes $\mu$, $\nu,\ldots$ run from 0 to
2 and Latin indexes from the beginning of the alphabet $a,\ b,\ldots$
(space indices)
take values 1, 2 or, later on, $x$, $y$. Three dimensional Lorentz 
frame indexes are denoted
by Latin capital letters $I,J,\ldots$ running from 0 to 2. Our
convention for the  tangent space-time metric is
$\eta_{IJ}=\mathrm{diag}(\sigma,1,1)$,
where $\sigma=\pm1$ allows us to switch between the Euclidean and
Lorentzian cases, respectively. The indices $I,J$, $\cdots$ are raised
and lowered with the metric $\eta_{IJ}$.}. The field equations read $F_{\Omega}=0$, where 
$F_{\Omega}\equiv\mathrm{d}\Omega+\Omega\, \Omega$
is  the field strength 2-form; this means that CS theory
is a topological theory with no truly propagating degrees of freedom.
By construction, the CS action is diffeomorphism invariant.

Next we assume $\mathcal{M}$ to be of topology  
$\Sigma\times\mathbb{R}$,
where $\Sigma$ is a two-dimensional manifold representing physical
2d space, and $\mathbb{R}$ representing time. The Hamiltonian formalism
can be achieved by splitting the connection into its temporal and
spatial components: $\Omega=\Omega_{t}\mathrm{d}t+\Omega_{a}\mathrm{d}x^{a}$.
Replacing this into (\ref{CS}) the action can be written as 
\begin{equation}
S=-\frac{\kappa}{2}\int_\mathbb{R}\mathrm{d}t\int_{\Sigma}
\langle\dot{
\boldsymbol{\Omega}},\boldsymbol{\Omega}\rangle
+2\langle\Omega_{t},\boldsymbol{F}_{\boldsymbol{\Omega}}\rangle,
\end{equation}
where $\boldsymbol{\Omega}=\OM_{a}\mathrm{d}x^{a}$ is the spatial
connection and $\boldsymbol{F}_{\boldsymbol{\Omega}}
=\boldsymbol{\boldsymbol{\mathrm{d}}}\boldsymbol{\Omega}+\boldsymbol{\Omega}\, \boldsymbol{\Omega}$
the associated spatial field strength.
The dot means time derivative.

Three-dimensional gravity meets CS theory when we choose as underlying
gauge symmetry the Poincar\'e group ISO(2,1). In this case, the connection
is written as 
\eq\ba{l}
\OM=e^{I}P_{I}+\omega^{I}J_{I} ,\esp
e^I=e^I_\m \mathrm{d}x^\m = \bfe^I+e^I_t \mathrm{d}t,\quad \bfe^I =e^I_a \mathrm{d}x^a,\esp
\om^I=\om^I_\m \mathrm{d}x^\m = \bfom^I+\om^I_t \mathrm{d}t ,\quad \bfom^I =\om^I_a \mathrm{d}x^a ,
\ea\eqn{CS-connection}
where $e$ and $\omega$ are the space-time 
 co-frame field and spin connection, $\bfe$ and $\bfom$ their spacial
counterparts, whereas $P_{I}$ and
$J_{I}$ correspond to the generators of translations and rotations
of ISO(2,1), respectively. We can go further and include a positive
(negative) cosmological constant $\Lambda$  deforming the gauge
symmetry to the (anti) de Sitter SO(3,1) (SO(2,2)) group. In any case,
the generators will satisfy the general Lie algebra given by $[J_{I},J_{J}]=\varepsilon_{IJ}{}^{K}J_{K}$,
$[J_{I},P_{J}]=\varepsilon_{IJ}{}^{K}P_{K}$ and 
$[P_{I},P_{J}]=\sigma\Lambda\varepsilon_{IJ}{}^{K}J_{K}$. 

A special feature of the SO(3,1) group is the possibility to define
two non-degenerate quadratic forms in the algebra. They correspond
to the two Casimir invariants 
\begin{equation}
\mathcal{C}_1=\eta^{IJ}P_{I}J_{J}\quad\quad\textrm{and}\quad\quad
\mathcal{C}_2=\eta^{IJ}(\frac{\sigma}{\Lambda} P_{I}P_{J}+J_{I}J_{J}).\label{Casimir:1}
\end{equation}
We can associate to each Casimir a corresponding inner product:
\begin{align*}
\langle P_{I},J_{J}\rangle_1 & =\eta_{IJ},\ \langle P_{I},P_{J}\rangle_1=0,\ 
\langle J_{I},J_{J}\rangle_1=0,
\end{align*}
\[
\langle P_{I},J_{J}\rangle_2 =0,\ \langle P_{I},P_{J}\rangle_2 =\sigma \Lambda \eta_{IJ},\ \langle J_{I},J_{J}\rangle_2 =\eta_{IJ}.
\]
The inner product defined by $\mathcal{C}_1$ is non-degenerate for
all $\Lambda$, whereas $\mathcal{C}_2 $ only for $\Lambda \neq0$.

Starting with the general action (\ref{CS}), we can write one action
for each inner product,
\begin{equation} 
\ba{l}
S_1  =-\dfrac{\kappa}{2}\dint_\mathbb{R} dt \dint_\S\eta_{IJ}
\left[\dot{\bfe}^{I}\boldsymbol{\omega}^{J}
+\dot{\boldsymbol{\omega}}^{I}\bfe^{J}+2e_{t}^{I}(\bfR
+\sigma \Lambda \mathsf{\bfe}^{2})^{J}
+2\omega_{t}^{I}\bfT^{J}\right],\esp
S_2   =-\dfrac{\kappa}{2}\dint_\mathbb{R} dt \dint_\S\eta_{IJ}
\left[\sigma \Lambda\dot{\bfe}^{I}\bfe^{J}
+\dot{\boldsymbol{\omega}}^{I}\boldsymbol{\omega}^{J}
+2\sigma \Lambda
e_{t}^{I}\bfT^{J}+2\omega_{t}^{I}(\bfR
+\sigma \Lambda \mathsf{\bfe}^{2})^{J}\right],
\label{CS:S}
\ea
 \end{equation} 
where  $\bfR^I=\mathbf{d}\bfom^I+\frac{1}{2}\e^I{}_{JK}\bfom^J\bfom^K$  is the spatial
curvature and $\bfT^I=\mathbf{d}\bfe^I+\e^I{}_{JK}\bfom^J\bfe^K$ the spatial torsion.
It can be recognized in $S_1$ the standard action
for 3d gravity. On the other hand, $S_2 $ can be considered
a kind of ``exotic'' 3d gravity in the sense
that, despite having a different action, it shares the same field
equations. It makes perfect sense to add the exotic action,
with an arbitrary coefficient $\gamma$, to the standard action, so
the most general action is 
\eq
S=S_1-\frac{1}{\gamma}S_2.
\eqn{general-action}
It is worth noticing that the equivalence can be established only
at the level of the equations of motions, but this is not true at the
level of the symplectic structure of the phase-space, as can be  seen from the
canonical Poisson brackets deduced from the kinetic part of the action, 
which will be displayed in Section \ref{lambda-positivo} after some
change of variables.
This intriguing model was studied in detail 
within the context of LQG  in~\cite{bonzom-livine}, 
where the appearance of $\gamma$ is compared
with the arbitrariness of the Barbero-Immirzi parameter~\cite{general-ref}.

This completes our brief review of Chern-Simons formulation of 3d
Gravity and the origin of the $\gamma$ parameter\footnote{Our notations slightly 
differ from those of \cite{bonzom-livine}. One recovers the latter from
ours by the substitutions $\sigma\to1$, $\g\to-\g/\sqrt{|\Lambda|}$ 
and $\kappa\to-2$,  where $\Lambda=s|\Lambda|$ is the cosmological 
constant with $s=-1,0,1$.}.

\section{Chern-Simons gravity with positive cosmological constant}
\label{lambda-positivo}

In what follows we will restrict the model to the $\Lambda>0$ sector. (The case of negative cosmological constant can be constructed analogously.) Also, we have kept open the possibility to switch between the Euclidean
and Lorentzian theories by introducing the parameter $\sigma=\pm1$,
so the gauge group would be $SO(4)$ or $SO(3,1)$, respectively.

Let us start by writing the connection as 
$\Omega=A^{i}L_{i}+B^{i}K_{i}$,
where $L_{i}$ are the generators of ``rotations'' and $K_{i}$ the generators
of ``boosts'' of the SO(3,1) (SO(4)) group, {\it i.e.}, in the SO(3,1) case, 
of its compact subgroup SO(3) and of its non-compact directions,
respectively.
These generators satisfy the Lie
algebra commutation rules 
\[
[L_{i},L_{j}]=\varepsilon_{ij}{}^{k}L_{k},\quad
[L_{i},K_{j}]=\varepsilon_{ij}{}^{k}K_{k},\quad
[K_{i},K_{j}]=\sigma\varepsilon_{ij}{}^{k}L_{k}. 
\]
Here $i,\; j,$
take values $1,\ 2,\ 3$ and are raised or lowered with the delta
Kronecker $\delta_{ij}$. Observe that the $A^{i}$ are recognized
as the components of an SO(3) (or SU(2)) connection.

The relations between the old and new generators and variables read
\begin{subequations}\label{dict}
\begin{align}
L & =(P_{2}/\sqrt{\Lambda},-P_{1}/\sqrt{\Lambda},\sigma J_{0}), 
& (A^i,\,i=1,2,3) & =(\sqrt{\Lambda}e^{2},-\sqrt{\Lambda}e^{1},
\sigma\omega^{0});\\
K & =(J_{2},-J_{1},P_{0}/\sqrt{\Lambda}), & (B^i,\,i=1,2,3) 
& =(\omega^{2},-\omega^{1},\sqrt{\Lambda}e^{0}).
\end{align}
\end{subequations}
As before, the non-degenerate invariant quadratic forms in the algebra are given
by the two Casimir of the group, but this time written in the new
basis\footnote{Since all group indices are contracted with the three dimensional
metric $\delta_{ij}$, it is convenient to adopt a vector-like notation,
\eg
$A^{i}B_{i}=A\cdot B$, 
$\e_{ijk}A^j B^k=(A\times B)_i$, etc.}:
 $\langle\Omega,\Omega'\rangle_1
=A\cdot B'+ B\cdot A'$
and $\langle\Omega,\Omega'\rangle_2 
=A\cdot A'+\sigma B\cdot B'$

With all this in mind, the general action \equ{general-action} can be written 
as\footnote{We recall that, here as in Eq. \equ{CS:S}, boldface letters
represent space objects (2-forms, etc.).}
 \begin{equation}
S=-\frac{\kappa}{2}\int_\mathbb{R}{\rm d}t\lp\int_\S
\left(\,\dot{\!\bfA}\cdot(\bfB-\frac{1}{\gamma}\bfA)+\dot{\bfB}\cdot(\bfA
-\frac{\sigma}{\gamma}\bfB)\right)-\mathcal{G}(A_{t})
-\mathcal{G}_{0}(B_{t})\rp,
\label{CS:SI-SII}\end{equation}
 where we are defining the smeared  quantities
\begin{subequations}
\begin{align}
\mathcal{G}(A_{t}) & =\kappa\int_\S A_{t}\cdot[{\bfD}\bfB
-\frac{1}{\gamma}(\bfF_{\bfA}+\frac{\sigma}{2}\bfB\times\bfB)],\\
\mathcal{G}_{0}(B_{t}) & =\kappa\int_\S B_{t}\cdot[\bfF_{\bfA}
+\frac{\sigma}{2}\bfB\times\bfB-\frac{\sigma}{\gamma}\bfD\bfB],
\label{GG_0}\end{align}
 \end{subequations} 
 together with $\bfF_{\bfA}=\mathbf{d}\bfA+\frac{1}{2}\bfA \times\bfA$ and
$\bfD\bfB=\mathbf{d}\bfB+\bfA\times\bfB$.

 One readily sees that the  theory defined by \equ{CS:SI-SII} is
fully constrained. The conjugate momenta $\Pi_i^{(A_t)}$ and
$\Pi_i^{(B_t)}$ of $A^i_t$ and $B^i_t$ are primary constraints, in
Dirac's terminology~\cite{dirac}, whereas $\GG(A_t)$ and $\GG_0(B_t)$
are the secondary constraints, $A^i_t$ and $B^i_t$ playing the role of
Lagrange multipliers. Other primary constraints involve the conjugate
momenta of the fields $A^i_a$ and $B^i_a$. They turn out to be of
second class, whose solution according to the Dirac-Bergmann
algorithm~\cite{dirac} gives rise to the Dirac-Poisson brackets
\begin{align}
\{A_{a}^{i}(\bfx),A_{b}^{j}(\bfx')\} & =
\frac{1}{\kappa}\e_{ab}\delta^{ij}\frac{\sigma\gamma}{\sigma-\gamma^{2}}
\delta^{2}(\bfx-\bfx'),\nonumber \\
\{B_{a}^{i}(\bfx),A_{b}^{j}(\bfx')\} & =
\frac{1}{\kappa}\e_{ab}\delta^{ij}\frac{\gamma^{2}}{\sigma-\gamma^{2}}
\delta^{2}(\bfx-\bfx')\label{Sym.2},\\
\{B_{a}^{i}(\bfx),B_{b}^{j}(\bfx')\} & =
\frac{1}{\kappa}\e_{ab}\delta^{ij}\frac{\gamma}{\sigma-\gamma^{2}}
\delta^{2}(\bfx-\bfx').\nonumber 
\end{align}
 From these relations we see that the 12 component of the fields 
$\bfA$ and $\bfB$ divide in 6 configuration fields and 6  momentum
fields.
It is important to mention that the gauge fields $\bfA$
and $\bfB$ are composed of mixed co-frame fields and spin connections
(see (\ref{dict})).

Performing the  Legendre transformation, we obtain a 
 fully constrained  
classical Hamiltonian, namely 
\begin{equation}
H=\mathcal{G} (A_{t})+\mathcal{G}_{0}(B_{t}),\label{Hamil}
\end{equation}
 with $A_{t}$ and $B_{t}$ Lagrange multipliers.
The algebra of the constraints closes under Poisson brackets, and adopts
the form:
\begin{align}
\{\mathcal{G}(\varepsilon),\mathcal{G}(\varepsilon')\} & 
=\mathcal{G}(\varepsilon\times\varepsilon').\nonumber\\
\{\mathcal{G}_{0}(\varepsilon),\mathcal{G}(\varepsilon')\} & 
=\mathcal{G}_{0}(\varepsilon\times\varepsilon'),\\
\{\mathcal{G}_{0}(\varepsilon),\mathcal{G}_{0}(\varepsilon')\} & 
=\sigma\mathcal{G}(\varepsilon\times\varepsilon')\nonumber. 
\end{align}
 We can recognize here the structure of the $\mathfrak{so}(3,1)$
($\mathfrak{so}(4)$) Lie algebra, in total agreement with the fact
that in the Dirac-Bergmann formalism for constrained systems, first
class constraints generate local gauge transformations. 
The infinitesimal gauge transformations generated by the constraints
are
\begin{equation}
\begin{aligned}
\{\mathcal{G}(\varepsilon),\bfA\} & =\bfD\varepsilon, & 
\{\mathcal{G}(\varepsilon),\bfB\} & = \bfB \times\varepsilon;\\
\{\mathcal{G}_{0}(\varepsilon'),\bfA\} & =\sigma\bfB\times\varepsilon' & 
\{\mathcal{G}_{0}(\varepsilon'),\bfB\} & = \bfD\varepsilon'
\quad\quad\quad (\bfD=\bfd+\bfA\times).
\end{aligned}
\label{Gau:Tr:BF}
\end{equation}
These gauge transformations are related, on-shell, with
local diffeomorphisms. This can be shown if we apply the Lie derivative
to the gauge fields
\begin{equation}
\begin{array}{cc}
\pounds_{\xi}\bfA & =\mathrm{D}(\imath_{\xi}\bfA)+\sigma(\imath_{\xi}\bfB)\times\bfB\ 
+\ \mbox{field equations,}\\
\pounds_{\xi}\bfB & =\mathrm{D}(\imath_{\xi}\bfB)+(\imath_{\xi}\bfA)\times\bfB\ 
+\ \mbox{field equations,}
\end{array}
\label{difeo}\end{equation}
with $\pounds=\mathrm{d}\imath_{\xi}+\imath_{\xi}\mathrm{d}$ the
Lie derivative. By comparison with (\ref{Gau:Tr:BF}) we  identify
in \equ{difeo}
infinitesimal gauge transformations with parameters 
$(\varepsilon,\varepsilon')=(\imath_{\xi}\bfA,\imath_{\xi}\bfB)$, 
up to field equations.

\section{Axial gauge}\label{calibre-axial}

We want to partially fix the gauge in such a way that the residual gauge
symmetry group be compact, namely  SO(3), in the SO(3,1) case to which we 
restrict hereafter. We shall verify that 
the gauge fixing condition $B^i_y\approx 0$, first is compatible with the Dirac
procedure (``Dirac compatible" according to the 
terminology of~\cite{bassetto}): it does indeed involve only the
phase space variables and leads to the presence of second class
constraints whose number is twice that of the gauge conditions, as we shall
verify. Second, it will reduce the gauge symmetry group in the desired
way.

To proceed with the gauge fixing, let us rewrite the
Hamiltonian\equ{Hamil},
adding the gauge condition as a constraint
\[
B^i_y(\bfx)\approx0 ,
\]
using a Lagrange multiplier field $\m(x)$:
\begin{equation}
H=\mathcal{G}(A_{t})+\mathcal{G}_0(B_{t}) 
+\dint_\S d^2x\,\mu_i(\bfx) B^i_{y}(\bfx) .
\end{equation}
 It turns out that the gauge fixing constraint together with the
constraint $\GG_0$ given  by  \equ{GG_0} are second class: indeed the matrix 
of their Poisson brackets,
\eq\ba{l}
C(\bfx,\bfy)=
\lp\ba{cc}
\{\GG_0^i(\bfx),\,\GG^j_0(\bfx')\}\ &\{\GG_0^i(\bfx),\,B^j_y(\bfx')\} \\
\{B^i_y(\bfx),\,\GG^j_0(\bfx')\}\ &\{B^i_y(\bfx),\,B^j_y(\bfx')\} 
\ea\rp\ES
\phantom{C(\bfx,\bfy)}\approx
\lp\ba{cc}
0\ &1\\
-1\ &0
\ea\rp
\lp\d_{ij}\pa_y +\e_{ijk}A_y^k\rp\d(\bfx-\bfx'),
\ea\eqn{matrix-secund}
is (weakly) non-singular.
Following the Dirac-Bergmann prescriptions, we introduce the Dirac
bracket
\[
\{M,\,N\}_{\rm D}= \{M,\,N\} - \{M,\,\chi_\a\} C^{-1}{}^{\a\b} \{\chi_\b,\,N\},
\]
where $M$ and $N$ are two phase space functions, 
$\chi_\a,\,\a=1,2$ are the two second class constraints and
$C^{-1}$ is the inverse -- in the convolution sense -- of the matrix
\equ{matrix-secund}. The Dirac brackets of the second class constraints with every phase
space function are zero by construction. Those of the fields $A^i_a$ are equal 
to their Poisson brackets \equ{Sym.2}:
\eq
\{A^i_{x}(\bfx),A^j_{y}(\bfx')\}_{\rm D} = 
\{A^i_{x}(\bfx),A^j_{y}(\bfx')\}  = 
 \dfrac{1}{\ka} \d^{ij}\dfrac{\s\g}{\s-\g^2}\d^2(\bfx-\bfx') .
\eqn{Dirac-brackets}
whereas those involving the remaining field $B^i_x$ are
different.
 We shall however not write down the latters, since this field
is not an independent variable. Indeed, the second class constraint are
now considered as strong equalities: in particular, the constraint
$\GG_0$ yields the equation
\[
\partial_{x}A_{y}-D_{y}\lp A_{x}-\frac{\sigma}{\gamma}B_{x}\rp =0 ,
\]
which can be solved  for $B_{x}^i$ as
a functional of $A^i_{x}$ and $A^i_{y}$.

At this stage, we are left with one set of first class constraints,
$\GG^i(\bfx)\approx0$.
Let us now define the new variables
\[
\mathcal{A}_{x} = A_{x}-\gamma B_{x} ,\quad
\mathcal{A}_{y} = A_{y} .
\]
Using the Dirac brackets \equ{Dirac-brackets} we can check that they
form a canonical pair of conjugate variables:
\eq
\{\mathcal{A}^i_x(\bfx),\,\mathcal{A}^j_y(\bfx')\}_{\rm D} =
\frac{\gamma}{\kappa}\d^{ij}\d^2(\bfx-\bfx') .
\eqn{DB-cal-A}
The Hamiltonian now reads
\eq
H=-\dfrac{\kappa}{\g}\GG(A_t) ,
\eqn{CS-hamiltonian}
with the first class constraint  $\GG$ given by
\eq
\GG(\eta) = \dint_\S d^2x\,\eta_i(\bfx)\FF^i\approx0 ,
\eqn{curv-constr}
where we have introduced the curvature  2-form associated to $\AA$:
\begin{equation}
\mathcal{F}^{i} =
\partial_{x}\mathcal{A}_{y}^{i}-\partial_{y}\mathcal{A}_{x}^{i}
+{\e}^i{}_{jk}\mathcal{A}_{x}^{j}\mathcal{A}_{y}^{k} .
\end{equation}
One recognizes in the Hamiltonian \equ{CS-hamiltonian} 
of the Chern-Simons theory for the
connection $\AA$, the latter tansforming as an  SO(3) or SU(2)
connection:
\[
\{\GG(\eta),\,\AA^i_a\}_{\rm D} = \pa_a\eta^i + \e^i{}_{jk}\AA^j_a \eta^k ,
\]
under the infinitesimal gauge transformations generated by the constraint
$\GG$.

Our gauge fixing has thus the effect of reducing the original gauge symmetry from
SO(3,1) to SO(3) or SU(2), so we end up with a CS theory with a compact
gauge group. This resembles the Ashtekar-Barbero variables formalism  in 3+1
dimensions. In fact, the new variables defined here can be considered
as the 2+1 analogous of the Ashtekar-Barbero variables. The role of the
present axial gauge can be compared with the temporal gauge. In both cases
the non-compact sector of the theory is frozen.

From this point, one can exploit several known approaches for quantizing
the CS theory with a SO(3) or SU(2) gauge group. For example, following the spirit
of the canonical quantization, we can mention the work of Dunne, Jackiw and
Trugenberger~\cite{quantiz-of-CS}, or~\cite{Constantinidis-Luchini-Piguet}
for a LQG inspired
treatment. The latter is summarized in the next section.

\section{ Quantization and Observables.
}\label{observavel}

\subsection{Quantum Theory}

In view of the result of the last section, 
we make here a quick review of the quantization of the
CS theory  with the gauge group $G=$ SU(2) on a time-oriented 3-manifold, 
$\mathcal{M}=\mathbb{R}\times\Sigma$.
This subsection essentially 
follows~\cite{Constantinidis-Luchini-Piguet}.

The canonical variables are defined as operators satisfying,
in correspondence with the Dirac brackets \equ{DB-cal-A},  the
commutation rules:
\begin{eqnarray}
[\hat{\mathcal{A}}_{x}^{i}(\bfx),\hat{\mathcal{A}}_{y}^{j}(\bfx')]
=\frac{i\gamma}{\kappa}\delta^{ij}\delta^{2}(\bfx-\bfx'),
\end{eqnarray}
where  $i,j=1,2,3$ are  the gauge group indices. 

We then  choose a polarization such that $\hat{\AA_{x}}$
is multiplicative, and $\hat{\mathcal{A}}_{y}$ is a functional 
derivative\footnote{ We use either the wave functional
representation or the abstract Dirac's kets. The relation between both
is given by $\Psi_\a[\AA_x]$ $=$ $\vev{\AA_x|\a}$, where $\a$ may
represent the quantum numbers defining the state.}
 acting on  wave functionals $\Psi[\mathcal{A}_{x}]$ 
$\equiv$ $\vev{\mathcal{A}|\Psi}$:
\begin{eqnarray}
\hat{\mathcal{A}}_{x}\Psi[\mathcal{A}_{x}] = \mathcal{A}_{x}\Psi[\mathcal{A}_{x}] & ; & 
\hat{\mathcal{A}}_{y}\Psi[\mathcal{A}_{x}] 
= \frac{\gamma}{i\kappa}\frac{\delta}{\delta
\mathcal{A}_{x}}\Psi[\mathcal{A}_{x}].
\end{eqnarray}
 In this  representation the Gauss constraint is written as
\begin{eqnarray}
\left[i\left(\partial_{x}\frac{\delta}{\delta \mathcal{A}_{x}^{i}}
+f^{i}{}_{jk}\mathcal{A}_{x}^{j}\frac{\delta}{\delta \mathcal{A}_{x}^{k}}\right)
+\dfrac{\kappa}{\g}\partial_{y}\mathcal{A}_{x}^{i}\right]\Psi[\mathcal{A}_{x}]
=0,
\label{gauss_c}\end{eqnarray}
and a particular solution is given by \cite{quantiz-of-CS} 
\eq
\Psi_{0}[A_{x}]=\exp(2\pi i\alpha_{0}),
\eqn{sol-Psi_0}
with 
\begin{eqnarray}
\alpha_{0}=\frac{\kappa}{6\pi\g}\int_{\tilde{\Sigma}}\epsilon^{\mu\nu\rho}{\rm Tr}(h^{-1}\partial_{\mu}h\, h^{-1}\partial_{\nu}h\, h^{-1}\partial_{\rho}h)d^{3}x-\frac{\kappa}{2\pi\g}\int_{ \Sigma=\partial {\tilde  \Sigma} }{\rm Tr}(\mathcal{A}_{x}h^{-1}\partial_{y}h),
\end{eqnarray}
The first term is the Wess-Zumino-Witten action,
and it is an integer  since the group is non-abelian and compact,  which requires
that $\kappa$ must be quantized,  $\kappa=\nu/4\pi,\, \n\in\mathbb{Z}$,
and $h\in G$ is defined  as a functional of $\mathcal{A}_{x}$ by 
\eq
\mathcal{A}_{x}=h^{-1}\partial_{x}h .
\eqn{def-h(A-thetat)}
It can be shown that taking into account the particular solution \equ{sol-Psi_0}, the
general wave functional solution of \equ{gauss_c} can be written as 
\eq
\Psi[\mathcal{A}_{x}]=\Psi_{0}[\mathcal{A}_{x}]\psi^{{\rm inv}}[\mathcal{A}_{x}],
\eqn{Psi-Psi0}
where $\psi{\rm ^{inv}}[\AA_{x}]$  satisfies
\begin{eqnarray}
\left[i\left(\partial_{x}\frac{\delta}{\delta \mathcal{A}_{x}^{i}}
+f^{i}\!_{jk}\mathcal{A}_{x}^{j}\frac{\delta}{\delta \mathcal{A}_{x}^{k}}\right)\right]
\psi^{{\rm inv}}[\mathcal{A}_{x}]=0.
\end{eqnarray}
The latter equation means that $\psi^{{\rm inv}}$ is invariant under the  
infinitesimal $x$-gauge  transformations
\begin{eqnarray}
\delta_{(x)} \mathcal{A}_{x}^{i}=D_{x}\epsilon^{i} .
\end{eqnarray}
At this point, one can choose to change  the focus from the functionals
 of the connection $\mathcal{A}_{x}$, which  does not transform homogeneously under the gauge 
transformations (as shown above), to functionals of the holonomies of $\mathcal{A}_{x}$ over 
a path $c_{y}=[x_{1},x_{2}]$
with constant $y$ on $\Sigma$, namely
\begin{eqnarray}
U(c_{y},x_{1},x_{2})=\mathcal{P}e^{\int_{x_{1}}^{x_{2}}\mathcal{A}_{x}^{i}(x,y) T_{i}dx},
\end{eqnarray}
whose transformations under the action of the gauge group are
\begin{equation}
U(c_{y},x_{1},x_{2})\mapsto g^{-1}(y,x_{2})U(c_{y},x_{1},x_{2})g(y,x_1).
\label{holotransf}
\end{equation}
Once these coordinates are themselves elements of the group, they turn out
to be more suitable for constructing a Hilbert space with a well-defined scalar product.
Taking into account   the considerations above we construct the cylindrical space Cyl  
  whose elements are  functionals (see \equ{Psi-Psi0}) 
\[\ba{l}
\Psi_{\Gamma,f}[\mathcal{A}_{x}] = 
\Psi_0[\mathcal{A}_{x}]\psi^{{\rm inv}}_{\Gamma,f}[\mathcal{A}_{x}],\esp
\mbox{with}\quad \psi^{{\rm inv}}_{\Gamma,f}[\mathcal{A}_{x}]
= f(U(c_{y_{1}},x_{1},x_{1}^{'}),...,U(c_{y_{K}},x_{K},x_{K}^{'})),
\ea\]
 where $\Gamma=\{c_{y_{k}},x_{k},x_{k}^{'},\, k=1,...,K\}$ is a graph
defined as a finite set of  $K$ $y$-constant paths on $\Sigma$,  and $f$ is a 
function on SU(2)$^{\times K}$ with complex values.
Such a state is denoted by $|\Gamma,f\rangle$.
Since the wave functionals are written in terms of a finite number
of holonomies, which are group elements,  the Haar measure  $dU_{k}$
may be used  to define the scalar product in Cyl: 
\begin{eqnarray}
\langle\Gamma,f|\Gamma,f'\rangle=\int\prod_{k=1}^{K}dU_{k}\overline{f(U_{1},...,U_{K})}f'(U_{1},...,U_{K}),
\end{eqnarray}
 The kinematical Hilbert space $\HH_{\rm kin}$ is then defined as 
the Cauchy completion of Cyl.
Making use of the Peter-Weyl theorem   one  finds a basis 
 $\ket{\GA,\vec j,\vec \a,\vec \b}$, with $\vec j=j_1,\cdots,j_K$, etc.
for $\HH_{\rm kin}$:
\begin{eqnarray}
\Psi_{\GA,\vec j,\vec \a,\vec \b} 
 & = & \Psi_{0}[\mathcal{A}_{x}]
\prod_{k=1}^{K} R_{\alpha_{k}}^{j_{k},\beta_{k}}(U(c_{y_{k}},x_{k},x'_{k})),
\end{eqnarray}
where $R_{\alpha}^{j,\beta}(h)$ denote the $(\alpha,\beta)$
matrix element of the spin $j$ representation of the holonomy.
Note that one excludes  the value $j=0$, and completes the basis with the ``null vector'' $\ket{0}$,
corresponding to the empty graph. These vectors form an orthogonal basis:
\begin{eqnarray}
\langle\Gamma,\vec{j},\vec{\alpha},\vec{\beta}|\Gamma',\vec{j'},\vec{\alpha'},\vec{\beta'}\rangle & = & 
 \d_{\GA\GA'}\delta_{\vec{j},\vec{j}'}\delta_{\vec{\alpha},\vec{\alpha}'}\delta_{\vec{\beta},\vec{\beta}'}
\end{eqnarray}
Thus, to every path $(c_{y_{k}},x_{k},x_{k}^{'})$ of the graph
$\Gamma$ we associate a spin $j_{k}$ representation of SU(2).
Vectors associated to different graphs are orthogonal.
Observe also that the kinematical Hilbert space is non-separable: 
it  is  the direct sum 
$\mathcal{H}_{\rm kin}=\bigoplus_{\Gamma}\mathcal{H}_{\Gamma}$, 
 over all graphs $\GA$, where $\mathcal{H}_{\Gamma}$ is the separable 
Hilbert space associated with the graph $\Gamma$.
 The expansion of a vector $\ket{\Psi}\in\HH_{\rm kin}$ reads
\[
\ket{\Psi} = \sum_{\GA,\vec{j},\vec{\alpha},\vec{\beta}}c_{\GA,\vec{j},\vec{\alpha},\vec{\beta}}
\ket{\GA,\vec j,\vec \a,\vec \b} ,
\]
where the sum over $\GA$ covers a countable  subset of graphs.

Before implementing the Gauss constraint \equ{gauss_c} we will specify the 
spatial two-dimensional manifold $\Sigma$ as 
being an infinite cylinder, and take $x$ as the periodic coordinate. 
This choice allows us to impose the 
constraint in the form of  the invariance of $\psi^{\rm inv}$
under all finite $x$-gauge transformations, 
implying, as can be easily
 realized from (\ref{holotransf}),
that  the functions should be reduced to  the traces of the holonomies along closed paths 
(cycles, or Wilson loops -- $y$-constant section of the cylinder) 
$U_{y}\equiv \textrm{Tr}\left( U(c_{y},x_1,x_1)\right)$, 
which depend on the $y$ coordinate, but not on $x$, after identifying the endpoints $x_1$ and $x_2$. 
Thus, each cycle is characterized by its ``height'', and the graphs are now sets $C$ of cycles.
This defines the Hilbert space $\mathcal{H}_{\rm Gauss}$, whose basis is 
the orthonormal set of ``spin network'' vectors $\vert C,\vec{j}\rangle$, given by
\begin{equation}
\begin{array}{ccc}
\Psi_{C,\vec{j}}[\mathcal{A}_{x}]
= \Psi_0[\mathcal{A}_x]\prod_{k=1}^{K}\chi^{j_{k}}(U_{y_{k}}) ,
& \quad &\textrm{with}\quad \chi^{j}(U_{y})=\textrm{Tr}\,R^{j}(U_{y}),
\end{array}
\end{equation}
where $\vec{j}$ stands for $(j_{1}\dots j_{K})$. These vectors are orthonormal, in the sense
\begin{equation}
\langle C,\vec{j}\vert C^{\prime},\vec{j}^{\prime}\rangle=\delta_{C,C^{\prime}}\delta_{\vec{j},\vec{j}^{\prime}} .
\end{equation}
Now we consider $S_{\circ}$, the space of all finite linear combinations 
of spin networks,  $\mathcal{H}_{\rm Gauss}$ being its Cauchy completion. 
It is the direct sum
 $\oplus_C\mathcal{H}_{\rm Gauss}^{C}$, where $\mathcal{H}_{\rm Gauss}^{C}$ 
is
the Hilbert space associated to a graph $C$, which is separable. This is not the case for 
$\mathcal{H}_{\rm Gauss}$, since the graphs are indexed by  finite arrays of real numbers.

Since $\vert C,\vec{j}\rangle$ depends on the $y$ coordinate, $\mathcal{H}_{\rm Gauss}$ 
is not diffeomorphism invariant. The  local invariance represented by 
the $y$-diffeomorphisms 
was not contemplated when we solved  the Gauss constraint, which must be corrected now. 
This is done with the use of the group averaging method (see \cite{general-ref}), 
based on the Gel'fand triple 
$S_{\circ} \subset \mathcal{H}_{\rm Gauss} \subset S^{\prime}_{\circ}$, 
being $S_{\circ}^{\prime}$ the dual of the spin-networks space $S_{\circ}$. 
The $y$-diffeomorphism invariant states are shown to be elements of this 
dual space constructed from any spin-network state through the application 
of a functional ``projector'' $P_{\rm diff}: 
S_{\circ}\rightarrow S_{\circ}^{\prime}$ defined by\footnote{ We use
Schwartz's notation $<\Phi,\,\Psi>$ for the value of the linear form
$\Phi\in S_{\circ}'$ applyed to the ``test'' vector $\Psi\in S_{\circ}$.}
\eq
\langle P_{\rm diff}\Psi,\Psi^{\prime}\rangle
=\sum_{\Psi^{\prime\prime}}\langle \Psi^{\prime\prime}
\vert \Psi^{\prime}\rangle, \quad \forall \vert 
\Psi^{\prime}\rangle \in S_{\circ},
\eqn{def-projector}
where the sum is done over all vectors 
$\vert \Psi^{\prime\prime}\rangle$ obtained from $\vert \Psi \rangle$ by 
a $y$-diffeomorphism.  The linear forms $\Phi=P_{\rm diff}\Psi$ 
span the physical Hilbert space $\mathcal{H}_{\rm phys}$,  with an interior product
induced from that of 
$\HH_{\rm Gauss}$~\cite{general-ref,Constantinidis-Luchini-Piguet}:
\eq
\vev{\Phi_1|\Phi_2} = \vev{\Phi_1,\Psi_2} = \vev{P_{\rm diff}\Psi_1,\Psi_2}.
\eqn{prod-int-phys}
The vectors of $\mathcal{H}_{\rm phys}$ only depend on the equivalence classes 
of spin-network states under $y$-diffeomorphisms. In particular, a state defined as explained 
above from $\vert C,\vec{j}\rangle$ does not depend on the particular positions $y_{k}$ of the cycles, 
but only on the number of such cycles and on the spin value associated with each of them. 
We have then  the s-knot states 
$\vert \vec{j}\rangle\equiv \vert j_{1},\dots,j_{K}\rangle=P_{\rm diff}\vert C,\vec{j}\rangle$,
which form an orthonormal basis for the physical Hilbert space. They are 
solution of the Gauss constraint and are invariant under all diffeomorphisms. 
Once the set of s-knots is countable, the physical Hilbert space is separable.

\subsection{ Observable}

Following the steps of what is done in Loop Quantum Gravity for defining 
the area operator,
we look for an operator  which be diagonal in the spin s-knot basis of 
the physical Hilbert space $\HH_{\rm phys}$.   

 We begin  with its construction in the Hilbert space 
$\HH_{\rm Gauss}$.
We start by defining an operator $\hat{W}_y$ such that it acts on the wave functionals 
 \equ{Psi-Psi0} in the following way:
\begin{equation}
\hat{W}^{i}_y\Psi[\mathcal{A}_{x}]=\Psi_{0}\hat{\mathcal{A}}_{y}^{i}\psi^{{\rm inv}}[\mathcal{A}_{x}] ,\label{eq:W_definition}
\end{equation}
To find the explicit form of this operator, let us assume we can split
it into two terms which basically separate the canonical variable dependence
from its conjugate momenta, \ie
\begin{eqnarray}
\hat{W}^{i}_y=\XX^{i}[\mathcal{A}_{x}]+\hat{\mathcal{A}}_{y}^{i} ,
\end{eqnarray}
with $\XX$ a    functional of the configuration variables to be defined.
Combining this with (\ref{eq:W_definition}), we get $\XX\Psi_{0}=-\hat{\AA}_{y}\Psi_{0}$.
 One shows easily~\cite{quantiz-of-CS} from the definition of $\Psi_0$  that 
$\hat{\AA}_{y}\Psi_{0}=(h^{-1}\partial_{y}h)\Psi_{0}$,
where $h=h[\AA_x]$ is the nonlocal functional  \equ{def-h(A-thetat)}. With this
result we can finally write:
\eq
\hat{W}_y=\hat{\AA}_{y}-h^{-1}\partial_{y}h.
\eqn{W_y}
The advantage to work with $\hat{W}^{i}_y$ is that this operator transforms as an SU(2) vector because it is defined as the difference of two objects that transform like connections, so it is a good 
 object with which we may construct gauge invariants observables.

 Let us calculate the action of $\hat{W}_{y}$ on the  particular spin $j$ 
wave functional
\eq
\Psi^{(j)}(\alpha,\AA) = \Psi_0[\AA_y] U^{(j)}(\alpha,\AA),\quad
 \mbox{with}\quad  U^{(j)}(\alpha,\AA) \equiv R^{(j)}(U(\alpha,\AA)),
\eqn{spinj-vector}
corresponding to  the holonomy along the $y$-constant 
path $x=\alpha(s)$, $s\in[0,1]$ on $\Sigma$,
in the spin $j$ representation. Once the result would be proportional to a 
$\delta$-distribution, it is more natural to consider  the action of the integrated  version
of $\hat{W}_{y}$ along a curve $y=\b(s)$ on $\Sigma$ at $x$ constant,
which we define as
\begin{equation}
\hat{W}(\beta)\equiv\int ds\dot{\beta}(s)\hat{W}_{y}(x,\beta(s)).
\end{equation}
Its action over the holonomy is  non-zero if and only if $\b$ and $\a$ intersect, with the result
\begin{equation}
\hat{W}^i(\beta ) \Psi^{(j)}= \frac{\gamma}{i\kappa} \Psi_{0}
U^{(j)}(\alpha_{1},\AA)T^{i(j)}U^{(j)}(\alpha_{2},\AA) ,\label{eq:A_betaU}
\end{equation}

where $\alpha_{1}$ and $\alpha_{2}$ are the parts of the curve $\a$ before and after its intersection
with $\b$, and $T^{i(j)}$ is  the representation matrix of the generator $T^i$. Observe that the action of 
$\hat{W}_y$ results in the insertion of this matrix  at the intersection point.
The quadratic operator 
$\hat{W}^2(\b)\equiv\sum_{i=1}^3\hat{W}^i(\b)\hat{W}^i(\b)$ 
acts on the same $\Psi^{(j)}$ as
\begin{equation}
\hat{W}^{2}(\beta)\Psi^{(j)}
=  \frac{\gamma^{2}}{\kappa^{2}}j(j+1) \Psi^{(j)},
\label{eigen-hol}\end{equation}
 where we have used the fact that the Casimir operator of  $SU(2)$ in the spin-$j$ representation is given by $\sum_{i=1}^3 T^{i(j)}T^{i(j)}=-j(j+1)\times\mathbf{1}^{(j)}$. 

In order to apply this operator to a general spin network vector, one needs 
to introduce a regularization. Similarly to the case of the area operator 
in $D=3+1$ 
LQG \cite{general-ref}, a regularization is available for the square root  
$\sqrt{\hat{ W}^2}$.
The graph $C$ of a spin network vector $\ket{C,\vec j}$ involves  
several cycles 
$C_n$, $n=1,\cdots N$, endowed with spin $j_n$ representations.
These cycles  cross the path \textbf{$\beta$} at different heights $y_n $. 
 Basically, the regularization scheme  consists in subdividing 
the path $\beta$
into $K$ segments $\beta_{k}$, $k=1,\cdots, K$, such that each 
cycle crosses at most
one of the segments $\beta_{k}$. For each of this segments we 
can compute the action of $\hat{L}_{k}$ defined as 
$\hat{L}_{k}\equiv\sqrt{\hat{W}^{2}(\beta_k)}$. 
Then the total 
regularized operator is the sum of all pieces: 
\eq
\hat{L}(\b)=\sum_{k=1}^{K}\hat{L}_{k}.
\eqn{reg-sum} 
Thus, when acting over a spin network vector, using the result 
\equ{eigen-hol} we obtain
\begin{equation}
\hat{L}(\b)\ket{C,\vec j} = L_{C,\vec j}(\b) \ket{C,\vec j},\quad
\mbox{with}\  L_{C,\vec j}(\b) 
= \frac{\gamma}{\kappa}\sum_{m=1}^{M}\sqrt{j_{m}(j_{m}+1)},
\label{L_general}\end{equation}
where the summation runs over all intersections of the curve $\b$ with
the graph $C$. Note that the eigenvalue $L_{C,\vec j}(\b)$ thus depends on
the graph $C$.
The result is independent of $K$, \ie of the refinement of 
the regularization scheme: it
only depends on the number $M$ of cycles crossing the path $\beta$ and on
their associated SU(2) representations. 
This defines a ``partial observable''~\cite{general-ref}, \ie
a self-adjoint  operator in $\HH_{\rm Gauss}$. 

An obvious question is about the possibility of extending the definition of 
this partial observable 
to an observable, \ie a self-adjoint operator  
$\hat L_{\rm phys}(\b)$ in $\HH_{\rm phys}$.
For any given finite curve $\b$, the answer is no, as we shall see now.
A natural extension of $\hat L(\b)$ as an operator acting in $S_\circ'$,
with domain in $\HH_{\rm phys}$, is
given by
\eq
\vev{{\hat L}'(\b)\Phi,\, \Psi'} 
= \vev{\Phi,\, \hat L(\b)\Psi'},\quad \forall \Psi'\in
S_\circ,
\eqn{def-L'}
where $\Phi\in \HH_{\rm phys}$, with  $\Phi=P_{\rm diff} \Psi$ for some
vector $\Psi\in S_\circ$ according to the definition 
\equ{def-projector}.
By linearity, it is sufficient to specialize to elements of the spin network basis of
$S_\circ$: $\Phi\to\Phi_{\vec j}=P_{\rm diff} \Psi_{C,\vec j}$ for some graph $C$, and
$\Psi'\to\Psi_{C',\vec j'}$.
We have 
\eq
\vev{\Phi_{\vec j},\,\Psi_{C',\vec{j}'}} 
= \dsum{C''\in[C]}{}\vev{\Psi_{C'',\vec j}|\Psi_{C',\vec{j}'}}
= \lac\ba{ll}  \d_{\vec{j}\vec{j}'} \ &\mbox{if}\ C'\in[C],\\
               0 &\mbox{if}\ C'\notin[C],
\ea\right.
\eqn{s-knotxspinnetwork}
where we have denoted by $[C]$ the $y$-diffeomorphism equivalence
class of the graph $C$. The last equality follows from the
orthonormality of the spin network basis. Then, using \equ{def-L'}:
\eq\ba{l}
\vev{{\hat L}'(\b)\Phi_{\vec j},\, \Psi_{C',\vec j'}}
= \vev{\Phi_{\vec j},\, \hat L(\b)\Psi_{C',\vec j'}}
= \dsum{C''\in[C]}{}\vev{\Psi_{C'',\vec j}|\hat L(\b)\Psi_{C',\vec{j}'}}\esp
= L_{C',\vec j}(\b) \dsum{C''\in[C]}{}\vev{\Psi_{C'',\vec j}|\Psi_{C',\vec{j}'}}
= \lac\ba{ll} L_{C',\vec j}(\b) \d_{\vec{j}\vec{j}'} \ &\mbox{if}\ C'\in[C],\\
               0 &\mbox{if}\ C'\notin[C],
\ea\right.
\ea\eqn{matrix-el-L}
where we have used the eigenvalue equation \equ{L_general}. We remark
that, since $L_{C',\vec j}$ depends on the graph $C'$ of the  spin
network vector $\ket{C',\vec j}$, argument of the linear form 
$\Phi_{\vec j}$, the s-knot vector $\ket{\vec j}$ is not an eigenvector
of ${\hat L}'(\b)$. Even more, the result, which should represent the
matrix elements of ${\hat L}'(\b)$ in the s-knot basis (see \equ{prod-int-phys}), 
is not even $y$-diffeomorphism invariant.

Let us show however that if, instead of the finite curves $\b$, we consider
infinite  curves $\b_\infty = \{x,y|x=\mbox{constant},-\infty<y<+\infty\}$,
we arrive at a well
defined observable, \ie a
self-adjoint operator $\hat L_{\rm phys}$ in $\HH_{\rm phys}$.
Indeed, first, Eq. \equ{L_general} becomes
\begin{equation}
\hat{L}(\b_\infty)\ket{C,\vec j} = L_{\vec j} \ket{C,\vec j},\quad
\mbox{with}\  L_{\vec j} 
= \frac{\gamma}{\kappa}\sum_{n=1}^{N}\sqrt{j_{n}(j_{n}+1)},
\label{L_general_infinite}\end{equation}
where the summation runs over all intersections of the  curve $\b_\infty$ with
the graph $C$, thus leaving an eigenvalue $L_{\vec j}$ independent
of the graph $C$ -- provided that the latter consists of $N$ cycles,
with spin atributes $\vec j=(j_1,\cdots,j_N)$. The eigenvalue is also independent
of the position $x$ of the curve $\b_\infty$, of course.
Going to the level of the physical Hilbert space, we can rewrite 
\equ{matrix-el-L} as a definition of the physical operator:
\eq\ba{l}
\vev{\hat L_{\rm phys}\Phi_{\vec j},\, \Psi_{C',\vec j'}}
= \vev{\Phi_{\vec j},\, \hat L(\b_\infty)\Psi_{C',\vec j'}}
= \dsum{C''\in[C]}{}\vev{\Psi_{C'',\vec j}|\hat L(\b_\infty)\Psi_{C',\vec{j}'}}\esp
= L_{\vec j} \dsum{C''\in[C]}{}\vev{\Psi_{C'',\vec j}|\Psi_{C',\vec{j}'}}
= \lac\ba{ll} L_{\vec j} \d_{\vec{j}\vec{j}'} \ &\mbox{if}\ C'\in[C],\\
               0 &\mbox{if}\ C'\notin[C],
\ea\right.
\ea\eqn{phys-matrix-el-L}
where $\hat L_{\rm phys}$ is defined from ${\hat L}'(\b)$ according to
\equ{def-L'} with obvious substitutions. Comparing with 
\equ{s-knotxspinnetwork}, we get the eigenvalue equations
\eq
\hat L_{\rm phys} \ket{\vec j} = L_{\vec j}  \ket{\vec j},
\eqn{L-phys-diagonal}
with $L_{\vec j}$ given by \equ{L_general_infinite}, which shows
that $\hat L_{\rm phys}$ is real diagonal in the s-knots basis of
$\HH_{\rm phys}$, hence defines an observable as announced.

When we restore the full dimensional parameters of the model we obtain
\begin{equation}
\kappa=\frac{c^{3}}{16\pi G\sqrt{\Lambda}}.
\end{equation}
Therefore \equ{L-phys-diagonal} reads
\begin{equation}
\hat{L}_{\rm phys}\,\ket{\vec j}
=\gamma16\pi
l_{\rm P}\sqrt{\Lambda}\sum_{n=1}^{N}\sqrt{j_{n}(j_{n}+1)}\,\ket{\vec j},
\end{equation}
where $l_{\rm P}=\hbar G/c^{3}$ is the Planck length. We can see in these
result an important difference which is the fact that the cosmological
constant appears in the formula.  
Because of this, $\hat{L}_{\rm phys}(\b)$ is dimensionless, 
but an operator with dimension of length can be defined as 
$\hat{L}_{\rm phys}/\sqrt{\Lambda}$ which has  a spectrum very similar 
to that of the area operator in $D=3+1$ gravity\footnote{ With the 
difference that in $D=3+1$ gravity the area operator is only a partial observable,
not even invariant under the space diffeomorphisms~\cite{general-ref}}.

We mentioned before that in Chern-Simons theory the \emph{level} of the
theory (i.e. the constant in front of the CS action) is quantized
by topological arguments when the underlaying gauge group is compact
and simply connected, then we have $\kappa/\gamma=\n/(4\pi\hbar)$,
with $\n\in\mathbb{Z}$. Therefore the last equation can be rewritten as
\begin{equation}
\hat{L}_{\rm phys}\,\ket{\vec j} 
= \frac{4\pi}{\n}\sum_{n=1}^{N}\sqrt{j_{n}(j_{n}+1)}\,\ket{\vec j}.
\end{equation}
This quantization rule can also be interpreted as a quantization rule of the
three fundamental constants of the theory:
\eq
4\g\sqrt{\LA} l_{\rm P}=\frac{1}{\n},\quad \n\in \mathbb{Z},  \quad \n\not=0.
\eqn{quant-fund=const}

\subsection{Classical Limit of the Observable $\hat L_{\rm phys}$}

The obvious question is which classical object corresponds to
$\hat{L}(\b)$ or $\hat{L}(\b_\infty)$. 
It is clear that the limit $K\to\infty$ of the classical 
counterpart of the sum \equ{reg-sum} is the Riemann sum of of the integral

\eq
L(\b) = \int ds\dot{\beta}\sqrt{\sum_{i=1}^3W_y^{i}W_y^{i}}
=\int ds\dot{\beta}\sqrt{-2\mathrm{Tr}(\AA_{y}-h^{-1}\partial_{y}h)^{2}},
\eqn{integral-L}

which reinforces the analogy with the $D=3+1$ area operator.

 We are going now to show that, for a finite curve $\beta$, and hence for
the curve $\beta_\infty$ in the limit, the classical gauge invariant
quantity $\hat{L}(\b)$ is vanishing.
In order to show this, let us first rewrite the CS connection 
$\AA=\AA_a \mathrm{d}x^a$ as
\[
\boldsymbol{\AA} = \LA^{-1}\bfd\LA + (\AA_x-\LA^{-1}\pa_x\LA)\mathrm{d}x, 
\]
where $\LA$ is defined as a functional of $\AA_y$ by the equation
$\LA^{-1}\pa_y\LA$ $=$ $\AA_y$. This shows that the connection is gauge
equivalent to $\boldsymbol{\AA}=E dx$, where 
$E=\LA(\AA_x-\LA^{-1}\pa_x\LA)\LA^{-1}$.
Aplying now the classical Gauss constraint \equ{curv-constr}, we obtain 
that the function $E$ is independent of $y$, and thus the connection is
gauge equivalent to 
\eq
\boldsymbol{\AA} = E(x)\mathrm{d}x.
\eqn{A=Edx'}
As a corollary, in the gauge where the latter equation holds, we have 
$\AA_y$ $=0$.
From its definition \equ{def-h(A-thetat)} as a functional of $\AA_x$,
we infer that $h[\mathcal{A}_x]$ is independent of $y$ in this gauge. 
Then $W_y$, as defined by the classical version of
\equ{W_y}, is vanishing. Since the classical quantity $L(\b)$ defined by
\equ{integral-L} is gauge invariant, we finally conclude that $L(\b)=0$,
hence $L(\b_\infty)=0$ in the limit, as announced. We are thus lead 
to the conclusion that the non-triviality of
the quantum observable $\hat L_{\rm phys}(\b)$ is a purely
quantum effect.

\noindent{\bf Remark.} That the gauge invariant local object $L(\b)=0$
be vanishing should be expected, since a topological theory is
chacterized by the absence of local invariant observables. 
In the limit of the curve $\b$ going to infinity, follows the vanishing 
of $L(\b_\infty)$,  although the latter is in fact global\footnote{It could a
priory depend on the position $x$ of the curve $\b_\infty$, but the
vanishing of the curvature  $\FF_{ab}$ and the non-Abelian Stoke 
theorem~\cite{stoke} allows to show easily that $L(\b_\infty)$ is
independent of $x$.}.

\section{Concluding remarks.}

The origin of the Barbero-Immirzi ambiguity $\gamma$ in Chern-Simons
formulation of 3d Gravity lies in the fact that it is possible
to define two non-equivalent inner products in the algebra of the $SO(3,1)$
group. Let's notice, however, that in 4d gravity the gauge group  is
usually taken as that of local Lorentz transformations, whereas in 
this case of 3d gravity, it is that of local Lorentz transformations
and translations (Poincar\'e, de Sitter or anti-de Sitter group). 
Therefore, it is expected a qualitative departure in 
the interpretation of $\gamma$ in both contexts -- 
although both gauge groups 
have the same dimension, namely six.

We have elaborated a detailed analysis of the $\Lambda>0$ sector of the
theory, where we have succeeded to reduce the de Sitter SO(3,1) gauge group to its 
{\it compact}
subgroup SO(3) thanks to a suitable axial gauge fixing. 
  We note that our results also apply to Riemannian gravity with 
negative cosmological constant, where the gauge group is de Sitter SO(3,1), too.
The case of Lorentzian gravity with  $\Lambda<0$ is not quite different, 
but technical difficulties 
to quantize the model are expected because  the $SO(2,2)$ group would 
only be reduced, by the same procedure, to a {\it non-compact} subgroup SO(2,1).

 We have thus obtained a notable reduction of 
the model. Specifically, in the positive $\LA$ case, we have got 
a Chern-Simons theory with an SU(2) gauge group. This result, in
particular the compactness of the residual gauge group, opened the way to
the LQG quantization  of a $D=2+1$ Lorentzian gravity theory.

Specializing to the particular case of a 2-dimensional space
manifold with the topology of a cylinder, we have applied a recent LQG
quantization scheme of the resulting CS 
theory~\cite{Constantinidis-Luchini-Piguet} which yields a physical
separable Hilbert space with a s-knot basis labeled by spin arrays 
$\vec{j}=(j_1,\cdots,j_N)$. We have constructed a global observable
$\hat{L}_{\rm phys}$, diagonal in that basis, with a spectrum very
similar to that of the partial observable ``area'' in $D=3+1$ gravity.

Finally, despite of the non-triviality of the quantum observable 
$\hat{L}_{\rm phys}$, we have found that its classical counterpart
$L(\b_\infty)$ is trivial, indeed null.  We have related this result to
the fact that a non-zero $L(\b_\infty)$ would be the limit of a local gauge invariant
quantity, hence of a classical local observable, $L(\b)$, and the existence of 
the latter would be in contradiction with the topological nature of the
theory.
Thus the observable $\hat{L}_{\rm phys}$ together with its spectrum 
appear as a purely quantum effect.


\small
\begin{thebibliography}{99}

\bibitem{Achucarro:1987vz}
A.~Achucarro and P.~K. Townsend,
{``A Chern-Simons Action for Three-Dimensional anti-De Sitter
  Supergravity Theories''},
{\em Phys. Lett.}, B180:89, 1986.

\bibitem{Witten:1988hc}
Edward Witten.
{``(2+1)-Dimensional Gravity as an Exactly Soluble System''},
{\em Nucl. Phys.}, B311:46, 1988.

\bibitem{Matschull:1999he}
Hans-Jurgen Matschull,
{``On the relation between 2+1 Einstein gravity and Chern- Simons
  theory''},
{\em Class. Quant. Grav.}, 16:2599--2609, 1999.

\bibitem{general-ref} C. Rovelli, ``Quantum Gravity'', Cambridge Monography on Math.
	Physics  (2004);\\
A. Ashtekar and J. Lewandowski, ``Background independent quantum gravity: 
        A status report'', \cqg{21}{2004}{R53}) [arXiv:gr-qc/0404018];\\
T. Thiemann,
 	``Modern Canonical Quantum General Relativity'',
   Cambridge Monographs on Mathematical Physics (2008);\\
M. Han, W. Huang and Y. Ma ``Fundamental structure of loop quantum  gravity'',
   \ijmp{D16}{2007}{1397}, [arXiv:gr-qc/0509064]; \\

\bibitem{barbero-immirzi} J.F. Barbero,  ``Reality conditions and Ashtekar
variables: A Different perspective'', Phys. Rev. D51 (1995) 5507, 
[arXiv: gr-qc/9410013]; \\
     Giorgio Immirzi, ``Real and complex connections for canonical
gravity'',\\ Class.Quant.Grav. 14 (1997) L177-L181, 
[arXiv: gr-qc/9612030]. 

\bibitem{bonzom-livine} V. Bonzom and E.R. Livine, 
``A Immirzi-like parameter for 3d quantum gravity''
\cqg{25}{2008}{195024}, [arXiv:0801.4241[gr-qc]]

\bibitem{ashtekar} A. Ashtekar, ``Lectures on Non-perturbative Quantum
gravity'', Notes prepared in collaboration with R. S. Tate, World Scientific, 
Singapore (1991).

\bibitem{nicolai-etc} H. Nicolai, K. Peeters and M. Zamaklar, 
   ``Loop quantum gravity: An outside view'', 
   \cqg{22}{2005}{R193},
	 [arXiv:hep-th/0501114];\\ 
H. Nicolai and K. Peeters, ``Loop and spin foam quantum gravity: A brief
   guide for beginners'', 	 [arXiv:gr-qc/0601129]; \\
T. Thiemann, 	``Loop quantum gravity: An inside view'', 
[arXiv:hep-th/0608210].

\bibitem{samuel} J. Samuel, ``A Lagrangian basis for Ashtekar's formulation of 
canonical gravity, {\em Pramana} 28 (1987) L429-L432

\bibitem{jacobson-smolin} Ted Jacobson and Lee Smolin,
``Covariant Action for Ashtekar's Form of Canonical Gravity'',
\cqg{5}{1988}{583}.

\bibitem{holst} S. Holst, ``Barbero's Hamiltonian derived from a generalized 
Hilbert-Palatini action'', Phys.Rev. D53 (1996) 5966-5969, 
[arXiv: gr-qc/9511026].

\bibitem{Constantinidis-Luchini-Piguet} C.P. Constantinidis, G. Luchini and O. Piguet,
``The Hilbert space of Chern-Simons theory on the cylinder.
A Loop Quantum Gravity approach'', \cqg{27}{2010}{065009},
[arXiv:0907.3240[gr-qc]].

\bibitem{thiemann-D>4} N. Bodendorfer, T. Thiemann and A. Thurn
``New Variables for Classical and Quantum Gravity in all Dimensions
I,II,III and IV''.

\bibitem{dirac}   P.A.M. Dirac, ``Lectures on Quantum Mechanics'',
Dover, 2001;   \\
M. Henneaux, C. Teitelboim, ``Quantization of Gauge Systems'',
Princeton University Press, 1994.

\bibitem{bassetto} A. Bassetto, G. Nardelli and R. Soldati, 
``Yang-Mills Theories in Algebraic Non-Covariant Gauges:
Canonical Quantization and Renormalization'', World Scientific, 1991.

\bibitem{quantiz-of-CS} G.V. Dunne, R. Jackiw, C.A. Trugenberger,
  ``Chern-Simons Theory in the Schr\"o\-din\-ger Representation'',
  \annp{194}{1989}{197}; \\
E.Guadagnini, M.Martellini, M.Mintchev, ``Braids and Quantum Group Symmetry in 
Chern-Simons Theory'', \np{B336}{1990}{581};\\
Steven Carlip, ``Quantum Gravity in 2+1 Dimensions'', 
Cambridge Monographs on Mathematical Physics (2003).

\bibitem{stoke} I.Ya. Arefeva, ``Non-Abelian Stokes Formula'',
{\em Theor. Math. Phys.}  43\,(1980)\,353;\\
Minoru Hirayama and Shugo Matsubara, ``Stokes Theorem for Loop Variables 
of Non-Abelian Gauge Field'', \ptp{99}{1998}{691}.

\end{thebibliography}
%


\end{document}